\documentclass[twocolumn,floats,floatfix,showpacs,amssymb,prd,superscriptaddress,nofootinbib]{revtex4-1}
\usepackage{graphicx, epsfig, amssymb} 
\usepackage{amsmath, amsfonts}
\usepackage{bm} 

\usepackage[linktocpage]{hyperref}
\usepackage[caption=false]{subfig}
\usepackage[usenames]{color}

\def\be{\begin{equation}}
\def\ee{\end{equation}}
\def\beq{\begin{eqnarray}}
\def\eeq{\end{eqnarray}}

\newcommand{\bea}{\begin{eqnarray}}
\newcommand{\eea}{\end{eqnarray}}
\newcommand{\ben}{\begin{enumerate}}
\newcommand{\een}{\end{enumerate}}
\newcommand{\bi}{\begin{itemize}}
\newcommand{\ei}{\end{itemize}}

\newcommand{\secsetup}{II}
\newcommand{\secsims}{III}
\newcommand{\secresults}{IV}
\newcommand{\secconclusion}{V}

\begin{document}

\title{\large Extreme black hole simulations: \\collisions of unequal mass black holes and the point particle limit}

\author{Ulrich Sperhake} \email{sperhake@ieec.uab.es}
\affiliation{Institut de Ci\`encies de l'Espai (CSIC-IEEC), Facultat de Ci\`encies, Campus UAB, E-08193 Bellaterra, Spain}
\affiliation{CENTRA, Departamento de F\'{\i}sica, Instituto Superior T\'ecnico, Universidade T\'ecnica de Lisboa - UTL,
Av.~Rovisco Pais 1, 1049 Lisboa, Portugal}
\affiliation{California Institute of Technology, Pasadena, CA 91125, USA}

\author{Vitor Cardoso} 
\affiliation{CENTRA, Departamento de F\'{\i}sica, Instituto Superior T\'ecnico, Universidade T\'ecnica de Lisboa - UTL,
Av.~Rovisco Pais 1, 1049 Lisboa, Portugal}
\affiliation{Department of Physics and Astronomy, The University of Mississippi, University, MS 38677, USA}

\author{Christian D. Ott} 
\affiliation{California Institute of Technology, Pasadena, CA 91125, USA}

\author{Erik Schnetter} 
\affiliation{Perimeter Institute for Theoretical Physics, 31 Caroline
  St.\ N., Waterloo ON N2L 2Y5, Canada}
\affiliation{Department of Physics, University of Guelph, 50 Stone Road East,
  Guelph, ON N1G 2W1, Canada}
\affiliation{Center for Computation \& Technology, 216 Johnston Hall,
  Louisiana State University, Baton Rouge, LA 70803, USA}
\affiliation{Department of Physics \& Astronomy, 201 Nicholson Hall,
  Louisiana State University, Baton Rouge, LA 70803, USA}

\author{Helvi Witek}
\affiliation{CENTRA, Departamento de F\'{\i}sica,
Instituto Superior T\'ecnico, Universidade T\'ecnica de Lisboa - UTL,
Av.~Rovisco Pais 1, 1049 Lisboa, Portugal}

\date{\today} 

\begin{abstract}
Numerical relativity has seen incredible progress in the last years,
and is being applied with success to a variety of physical phenomena,
from gravitational-wave research and relativistic astrophysics
to cosmology and high-energy physics. Here we probe
the limits of current numerical setups, by studying collisions of unequal
mass, non-rotating black holes of mass-ratios up to 1:100 and making
contact with a classical calculation in General Relativity: the infall
of a point-like particle into a massive black hole.

Our results agree well with the predictions coming from linearized
calculations of the infall of point-like particles into non-rotating black
holes. In particular, in the limit that one hole is much smaller than
the other, and the infall starts from an infinite initial separation,
we recover the point-particle limit. Thus, numerical relativity is
able to bridge the gap between fully non-linear dynamics and linearized
approximations, which may have important applications. Finally, we also
comment on the ``spurious'' radiation content in the initial data and
the linearized predictions.
\end{abstract}
\pacs{04.25.D-, 04.25.dg, 04.25.Nx, 04.30.-w}

 
 


\maketitle

\section{\label{sec:intro} Introduction}
In recent decades, black holes have started playing a key role
in a variety of processes in astrophysics, gravitational wave
physics and high-energy physics.  Following the 2005 breakthroughs
\cite{Pretorius:2005gq,Campanelli:2005dd,Baker:2005vv},
numerical relativity has been an essential tool in the modeling
of black-hole binaries in the strong-field regime. At the same
time it has become clear that detailed studies of black-hole
systems often involve a close interplay between fully non-linear
numerical simulations and approximation techniques of various
types. For example, the generation of gravitational wave (GW)
template banks for use in the analysis of observational data from
laser interferometric GW detectors LIGO, VIRGO, GEO600, LCGT or LISA
requires the combination of numerical relativity with post-Newtonian
or other techniques; see Refs. \cite{Berti:2007fi,Centrella:2010mx,
Damour:2010zb,Pan:2009wj,Santamaria:2010yb,Abadie:2011kd,MacDonald:2011ne}
and references therein. Post-Newtonian studies have also played an
important role in the guidance of the numerical investigation of the
black-hole recoil, most notably in the discovery of the so-called
{\em superkicks} and its possible suppression due to spin alignment
\cite{Gonzalez:2007hi,Campanelli:2007cga,Campanelli:2007ew, Kesden:2010yp,
Kesden:2010ji}. In the context of high-energy collisions of black holes,
linearization tools such as the zero-frequency limit or point-particle
calculations provide valuable insight into the scattering threshold and
GW emission of black-hole collisions in four and higher-dimensional
spacetimes \cite{Berti:2010ce}.  A particular class of black-hole
binaries of high relevance for the spaceborne LISA (or a similar
future spaceborne) observatory, the
so-called extreme-mass-ratio inspirals, represent a particularly
difficult challenge to numerical relativity and their modeling
relies heavily on perturbative methods and self-force calculations; see
Refs. \cite{Hinderer:2008dm,Barack:2009ux,Canizares:2009ay,Yunes:2010zj,
Sundararajan:2010sr,Poisson:2011nh} and references therein.

With the above as motivation, it is vital to obtain a detailed
understanding of the range of validity of the various types of
approximation methods. At the same time, these methods provide valuable
tools to calibrate the accuracy of numerically generated solutions to the
Einstein equations. The purpose of this paper is to provide such a study
for the case of a classical calculation in general relativity, the head-on
infall of a point-particle (PP) into a black hole \cite{Davis:1971gg}.

In recent years, numerical relativity has started probing the
intermediate mass-ratio regime by evolving the final orbits of
(approximately) quasi-circular inspirals of black-hole binaries with
mass-ratio $q\equiv m_2/m_1=1/10$ \cite{Gonzalez:2008bi, Lousto:2010tb};
by comparing numerical results with perturbative calculations employing
the fully numerical black-hole trajectories for mass ratios up to $q=1/20$
\cite{Lousto:2010qx} and most recently, the first numerical evolution of
a black-hole binary with $q=1/100$ \cite{Lousto:2010ut}.  In this work,
we restrict our attention on the head-on limit of the collision of black
holes, for two reasons: (i) the lower computational cost due to the higher
degree of spacetime symmetry and the absence of the lengthy inspiral
phase and (ii) the availability of high-precision results in the PP limit.

In our study we will make extensive use of the calculation by Davis {\em
et al.} \cite{Davis:1971gg} who model in the PP limit the collision of
a small object of mass $m$ with a black hole of mass $M \gg m$.  In the
original calculation the particle was falling from rest at infinity,
and the total radiated energy was found to be
\begin{equation}
E^{\rm rad}_{\rm PP}=0.0104 \frac{m^2}{M}\,.\label{DRPP}
\end{equation}
This setting has been generalized to arbitrary initial distance and boost,
in which case initial data and consequent spurious radiation play a role
\cite{Lousto:2004pr,Lousto:1996sx,Cardoso:2002ay,Berti:2010ce,Mitsou:2010jv}.

Fully numerical results for black-hole head-on collisions obtained
in the equal and comparable mass regime have been compared with PP
predictions and results obtained in the close-limit approximation
\cite{Andrade1997} by Anninos and collaborators \cite{Anninos:1994gp,
Anninos:1998wt}. These studies demonstrated agreement for the radiated
energy and linear momentum bearing in mind the accuracies achievable at
the time. However, the spurious radiation present in the initial data
was not dealt with in a satisfactory manner. Its presence contaminated
the physical pulse and much of the conclusions in these earlier works
were affected by this. In particular, the presence of spurious radiation
prevented an accurate computation of the total radiation and comparison
with the linearized, PP calculations.

Also, we are not aware of any comparisons between PP calculations
and fully numerical results for mass ratios in a truly perturbative
regime. By simulating black-hole binaries up to a mass ratio of $q=1/100$
we fill this gap and identify those aspects of the PP predictions
which describe black-hole dynamics well in general and which only hold
in the extreme mass-ratio limit. From a different point of view, the
agreement with the PP calculations represents an important
validation of the fully numerical calculations in the regime of
high-mass ratios. In this context we emphasize that we are able to
accurately extract from binary black-hole simulations
radiated GW energies of the order of $10^{-6}~M$
and linear momenta corresponding to recoil velocities of a few dozens
of m/s, similar to the average speed of a normal car.
We note, however, that even smaller amounts of energy have been extracted from
general relativistic simulations of stellar core collapse;
see e.~g.~\cite{Reisswig2010a}.

This paper is organized as follows. We summarize our numerical
framework in Sec.~\secsetup, estimate numerical uncertainties in
Sec.~\secsims, describe our results in Sec.~\secresults\ and conclude
in Sec.~\secconclusion.

\vspace{0.2cm}
\section{Numerical Setup and Analysis Tools.}
The numerical simulations of unequal-mass black-hole collisions
starting from rest have been performed with the \textsc{Lean} code,
originally introduced in Ref.~\cite{Sperhake:2006cy,Sperhake:2007gu}.
The \textsc{Lean} code is based on the \texttt{Cactus} computational
toolkit \cite{Goodale02a, cactus} and uses the \texttt{Carpet} mesh refinement
package \cite{Schnetter:2003rb, carpet},
 the apparent horizon finder \texttt{AHFinderDirect}
\cite{Thornburg:1995cp,Thornburg:2003sf} and the \textsc{TwoPuncture}
initial data solver~\cite{Ansorg:2004ds}. The $3+1$ Einstein's
equations are evolved using the BSSN \cite{Shibata:1995we,
Baumgarte:1998te} formulation, together with the moving puncture
approach \cite{Baker:2005vv,Campanelli:2005dd}. The gauge conditions
are determined by the so-called puncture gauge, i.e., the ``1+log''
slicing and $\Gamma$ driver shift condition \cite{Alcubierre2003b}.
The systems are set up
using Brill-Lindquist initial data.
We have evolved BH binaries with mass ratios $q \equiv m_2/m_1
= 1, 1/2, 1/3, 1/4, 1/10$ and $1/100$, where $m_i$ is the bare mass parameter
of the $i$-th BH.

We use the Newman-Penrose scalar $\Psi_4$ to measure gravitational radiation
at extraction radii $R_{\rm ex}$, chosen in a range of $40~M$ to $90~M$
from the center of the collision. We decompose
$\Psi_4$ into multipoles $\psi_{lm}$ using spherical harmonics of
spin-weight $-2$, ${_{-2}}Y_{lm}$, according to
$rM\Psi_4(t,r,\theta,\phi)=\sum_{l=2}^\infty \sum_{m=-l}^l
\,{_{-2}}Y_{lm}(\theta\,,\phi)\, \psi_{lm}(t,r)$.
Due to the symmetry properties of the systems under consideration,
the only non-vanishing multipoles all have $m=0$ in a suitably chosen frame, and are purely real,
corresponding to a single polarization state $h_+$. In the equal-mass limit,
the additional symmetry causes all multipoles with odd $l$ to vanish
identically. The energy spectrum and luminosity of the radiation
are given by
\begin{eqnarray}
\frac{dE}{d\omega}&=&\sum_l
\frac{1}{16\pi^2}\frac{|\hat{\psi}_{l0}(\omega)|^2}{\omega^2}
\equiv \sum_l \frac{dE_l}{d\omega}\,,\label{spectrum} \\
\frac{dE}{dt}&=&\sum_l \frac{1}{16\pi M^2} \left|
\int_{-\infty}^{t} \psi_{l0}(\tilde{t}) d\tilde{t} \right|^2
\equiv \sum \frac{dE_l}{dt}\,,
\label{luminosity}
\end{eqnarray}
respectively, where a hat denotes the Fourier transform and $\psi_{l0}$
is evaluated on a sphere at
infinity.

\vspace{0.2cm}
\section{Simulations and uncertainties}
\label{sec:sims}
We have performed a series of simulations of head-on collisions with mass
ratio ranging from $q=1$ to $q=1/100$ with initial coordinate separation
$d$ and proper horizon-to-horizon separation $L$ as given in
Table~\ref{tab:models}.
\begin{table}
\begin{tabular}{r|rr|cccc|c}
  \hline
  $q$ & $d/M$ & $L/M$ & $E^{\rm rad}/M$ & \multicolumn{3}{c}{$E^{\rm rad}_{l=2,3,4}(\%)$} & $v/({\rm km}/{\rm s})$ \\
  \hline
    1 & 10.24 & 12.48 & $5.32 \times 10^{-4}$ & 99.6 & 0    & 0.03 & 0 \\
    1 & 12.74 & 16.76 & $5.39 \times 10^{-4}$ & 99.3 & 0    & 0.03 & 0 \\
    1 & 17.51 & 21.82 & $5.56 \times 10^{-4}$ & 99.4 & 0    & 0.03 & 0 \\
  1/2 & 12.74 & 16.69 & $4.33 \times 10^{-4}$ & 98.1 & 1.28 & 0.07 & 3.71 \\
  1/3 & 12.74 & 16.60 & $3.11 \times 10^{-4}$ & 96.7 & 2.83 & 0.16 & 3.97 \\
  1/4 &  7.31 & 10.57 & $2.16 \times 10^{-4}$ & 95.8 & 3.85 & 0.25 & 3.65 \\
  1/4 & 12.74 & 16.53 & $2.28 \times 10^{-4}$ & 95.4 & 4.14 & 0.28 & 3.72 \\
  1/4 & 17.51 & 21.61 & $2.33 \times 10^{-4}$ & 95.6 & 4.13 & 0.27 & 3.83 \\
 1/10 & 12.72 & 16.28 & $6.05 \times 10^{-5}$ & 92.1 & 7.09 & 0.67 & 1.31 \\
 1/10 & 16.72 & 20.55 & $6.16 \times 10^{-5}$ & 92.5 & 7.23 & 0.70 & 1.33 \\
 1/10 & 20.72 & 24.76 & $6.29 \times 10^{-5}$ & 92.0 & 7.15 & 0.67 & 1.34 \\
1/100 &  7.15 &  9.58 & $9.10 \times 10^{-7}$ & 88.1 & 9.01 & 1.15 & 0.0243 \\
1/100 & 11.87 & 15.08 & $9.65 \times 10^{-7}$ & 88.0 & 9.87 & 1.46 & 0.0248 \\
1/100 & 13.85 & 17.21 & $9.94 \times 10^{-7}$ & 87.8 &10.11 & 1.46 & 0.0256 \\
1/100 & 15.08 & 18.53 & $1.012 \times 10^{-6}$ & 87.7 &10.05 & 1.51 & 0.0260 \\
\hline
\end{tabular}
\caption{Mass ratio $q$, coordinate and proper separation $d$ and $L$,
respectively, as well as radiated energy $E_{\rm rad}$ with percentage
distribution in the $l=2$, $l=3$ and $l=4$ multipoles and recoil velocity
$v$ for the set of binary models evolved numerically.
}
\label{tab:models}
\end{table}
We describe the grid setup used for these simulations in terms of the
number $n_{\rm rl}$ of refinement levels, the
radius $R$ of the computational domain, the resolution $H$ used in the
wave extraction zone\footnote{Typically the third refinement level counted
from the outside.}, the radius $r$ in units of the smaller hole's mass $m_2$
of the innermost refinement level centered on the individual
punctures\footnote{For the small mass ratios $q=1/10$ $(1/100)$, the two (five)
highest resolution boxes are placed around the small hole only to reduce
computational cost.} and the resolution $h/m_2$ of the innermost
refinement level. The values for these parameters are summarized for all
mass ratios in Table~\ref{tab:grids}.
\begin{table}
\begin{tabular}{r|ccccc}
  \hline
  $q$ & $n_{\rm rl}$ & $R/M$ & $H/M$ & $r/m_2$ & $h/m_2$ \\
  \hline
    1 & 9 & 512 & 0.76 & 2 & 1/21 \\
  1/2 & 9 & 341 & 0.51 & 2 & 1/21 \\
  1/3 & 9 & 256 & 0.76 & 2 & 1/21 \\
  1/4 & 9 & 205 & (1.22,~1.07,~0.95) & 1 & (1/21,~1/24,~1/27) \\
 1/10 &12 & 303 & 0.73 & 0.625 & 1/64 \\
1/100 &15 & 223 & (1.01,~0.63,~0.51) & 0.625 & (1/40,~1/64,~1/80) \\
  \hline
\end{tabular}
\caption{Grid setup used for the different mass ratios $q$. The number
         of refinement levels is given by $n_{\rm rl}$,
         $R$ is the
         radius of the computational domain, $H$ the resolution in the
         wave extraction zone, $r$ the radius of the innermost refinement
         box around the individual punctures and $h$ the resolution
         used on that level. The additional low and high resolution for
         $q=1/4$ and $q=1/100$ have been used for the convergence studies.
}
\label{tab:grids}
\end{table}

Our results are affected by three main sources of uncertainties: finite
extraction radius, discretization and, for small initial separations of the
binary, spurious initial radiation. We reduce
the error arising from finite extraction radius by measuring the waveform
components at several radii, and fitting them to an expression of the form
$\psi_{lm}(r,t)=\psi_{lm}^{(0)}(t)+\psi_{lm}^{(1)}(t)/r$.
The waveform ``at infinity'' $\psi_{lm}^{(0)}(t)$ is the quantity reported
throughout this work and used to calculate related quantities, such as the
radiated energy. The uncertainty in this extrapolated value is estimated by
performing a second fit including also a quadratic term $\psi_{lm}^{(2)}/r^2$,
and taking the difference between the first- and second-order fits. The
resulting uncertainty increases as we decrease the mass ratio $q$ and
is $1-4~\%$ for the total radiated energy and the
$l=2$ waveform and energy, and $3-5~\%$ for the subdominant multipoles
and the radiated linear momentum.

In order to estimate the discretization error of our simulations, we
have performed a convergence analysis for models $(q=1/4,~L=16.53~M)$
and $(q=1/100,~L=9.58~M)$ using the three resolutions listed for
these mass ratios in Table~\ref{tab:grids}.
\begin{figure}
\includegraphics[clip=true,width=0.48\textwidth]{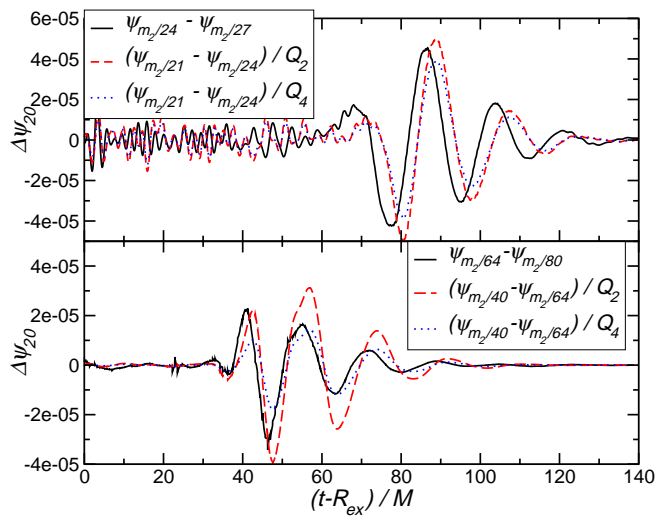}
\caption{Convergence analysis for the $l=2$ multipole of the gravitational
         wave signal for simulation $q=1/4,~D=16.53~M$ (upper) and
         simulation $q=1/100,~D=9.58$ (lower panel). In both cases we
         show the higher resolution differences (solid black)
         together with the lower
         resolution result rescaled for second (dashed red lines) and
         fourth-order convergence (dotted blue lines).
}
\label{fig:conv_psi20}
\end{figure}
The resulting convergence plots for the $l=2$ multipole of the wave signal
is shown in Fig.~\ref{fig:conv_psi20} and demonstrates convergence between
second and fourth order.
With regard to the analysis below, we note in particular that
the $q=1/100$ case exhibits second order convergence in the plunge-merger
signal around $t-R_{\rm ex} \approx 40~M$ but is close to fourth-order
convergence for the remainder of the waveform. Bearing in mind that
the plunge-merger transition represents the most dynamic part of the
evolution and that the second-order ingredients in the code are associated
with the prolongation of grid functions at the refinement
boundaries in time, this observation is
compatible with the numerical discretization. We observe similar
convergence properties for the $l=3$ multipole, but overall convergence
close to fourth-order for the radiated energy and linear momentum,
presumably because the accumulated errors are dominated by the
fourth-order contributions observed for most of the signal.
The resulting numerical uncertainties for $q=1/100$ are about $10~\%$
in the waveform for the plunge-merger transition and $5~\%$ for the
remainder of the signal as well as $6~\%$ for the radiated energy
and $8~\%$ for the linear momentum lost in gravitational waves.
We note that in both cases, the discretization error leads to
an overestimate of the radiated quantities.
For $q=1/4$ we observe significantly smaller uncertainties in the
range of $2~\%$ for all quantities.

Finally, we comment on the unphysical gravitational radiation inherent in
the conformally flat puncture initial data.
In order to extract physically meaningful information, one
has to separate the spurious radiation from the radiation generated by the
collision itself. This is done by ``waiting'' for the spurious radiation to
radiate off the computational domain, and then discarding the early,
contaminated part of the wave signal. For small values of the
initial separation, however, the binary will merge before the
spurious radiation has had enough time to leave the system, and physical
and unphysical contributions to the wave signal partially overlap
and cannot be cleanly distinguished. For our set of simulations, this
problem arises only in the case $q=1/100$, $L=9.58~M$, where it introduces
an additional error of about $2~\%$ to the radiated energy and momentum.

\vspace{0.2cm}
\section{Results}
All collisions summarized in Table \ref{tab:models} result in the formation
of a single BH plus gravitational radiation, i.~e.~there is no indication
of violation of the cosmic censorship conjecture. The final BH is born
distorted, and eventually rings down to a Schwarzschild solution via
emission of a superposition of quasi-normal modes \cite{Berti:2009kk}.
\begin{figure}
\begin{center}
\includegraphics[clip=true,width=262pt]{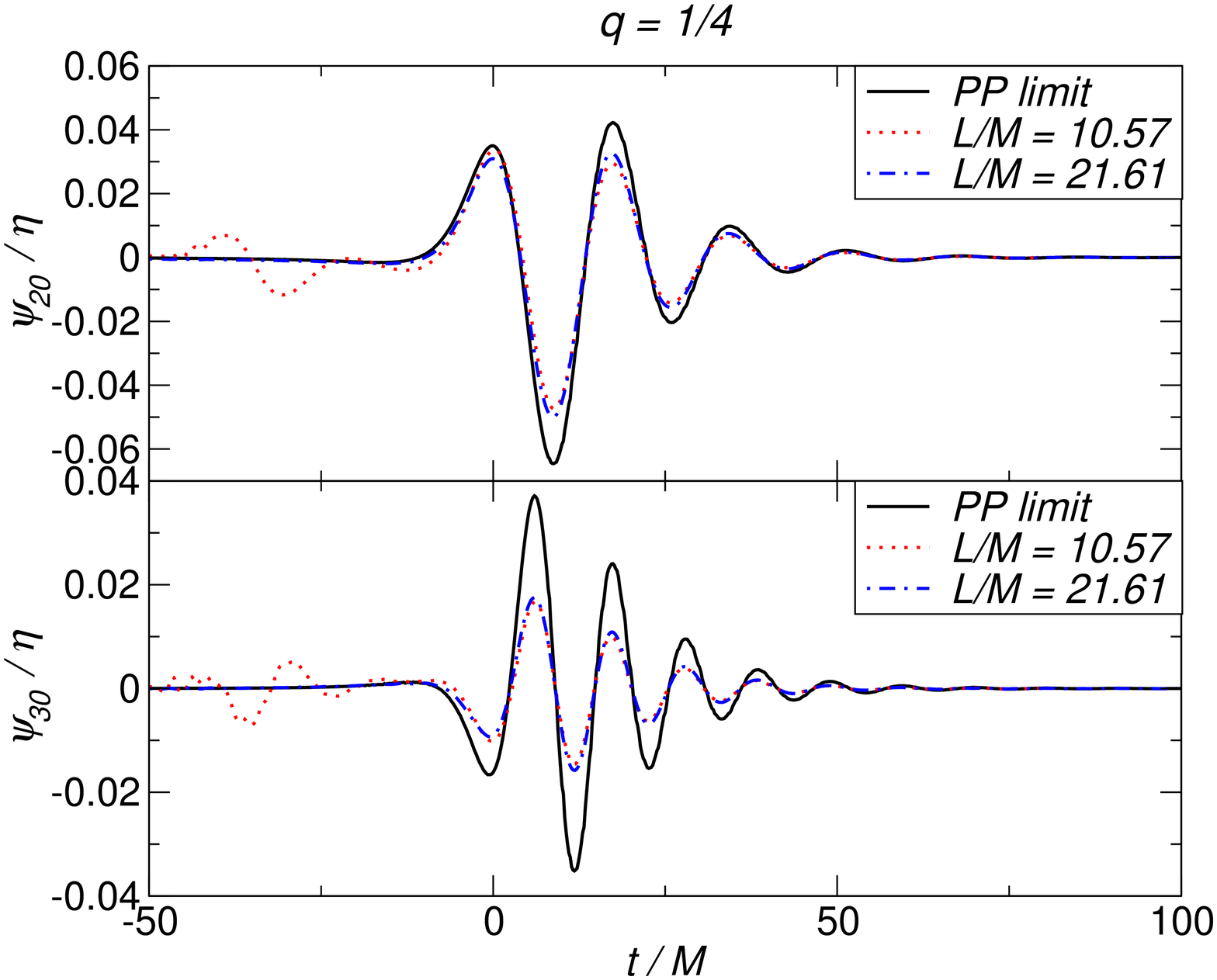}
\hspace{-0.85cm}
\includegraphics[clip=true,width=262pt]{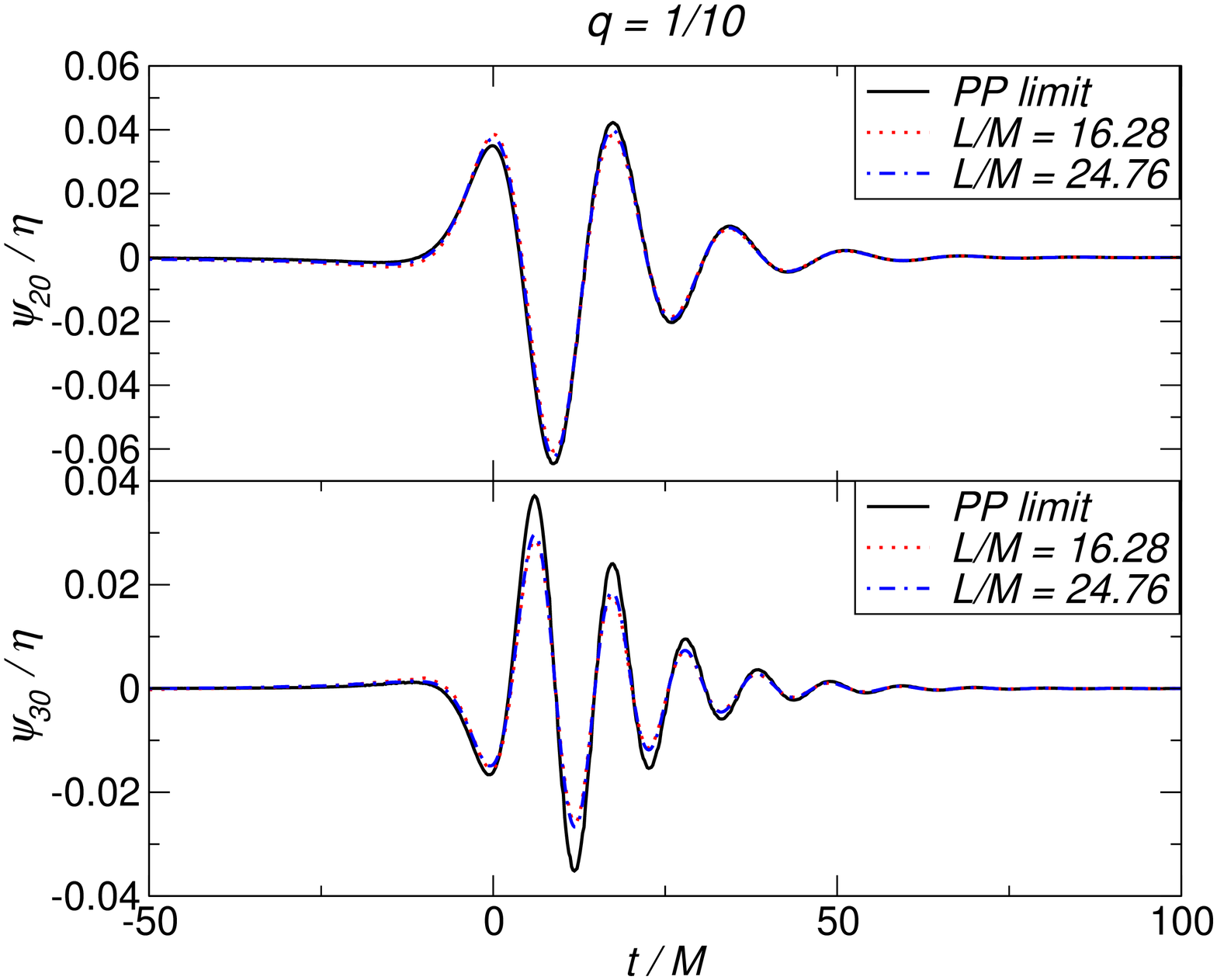}
\hspace{-0.85cm}
\includegraphics[clip=true,width=237pt]{f2c_PP_q100}
\end{center}
\caption{\label{fig:waveformsPP} (Color online)
Rescaled waveforms for mass ratios
$q=1/4$ (top), $q=1/10$ (center) and $q=1/100$ (bottom panels)
for $l=2$ (upper) and $l=3$ (lower half of each panel),
for two different initial separations. Also shown is the waveform in the 
PP limit (black solid lines).
}
\end{figure}

We illustrate the $l=2$ and $l=3$
wave signal in Fig.~\ref{fig:waveformsPP} for the
$l=2$ and $l=3$ multipoles obtained for the mass ratios $q=1/4$ (top),
$q=1/10$ (center) and $q=1/100$ (bottom).
In each panel the solid (black) curves represent
the PP prediction for infall from infinity whereas the
dotted (red) and dash-dotted (blue) curves show the numerical results
for different values of the finite initial separation.
To leading order, the gravitational radiation output of black-hole collisions
scales with the square of the reduced mass
$\mu\equiv M\eta$ of the system, where $\eta=q/(q+1)^2$ is
the dimensionless, symmetric mass ratio \cite{Davis:1971gg}.
For comparison of the numerical
results with PP predictions, we therefore rescale the
former by the corresponding powers of $\eta$, 
quadratic for energy and linear for the waveforms in Fig.~\ref{fig:waveformsPP}.

The waveforms show interesting features. For small initial separations,
the early part of the waveform is contaminated by ``spurious''
radiation; cf.~the dotted (red) curve in the top and bottom panels of
Fig.~\ref{fig:waveformsPP}. As the initial separation increases,
however, this problem disappears, because the longer infall duration of
the binary provides sufficient time for the unphysical radiation to propagate
off the grid; cf.~the dash-dotted (blue) curves. A closer inspection
of the $q=1/100$ case yields excellent agreement between the numerical
and PP predictions except for the plunge-merger transition
around $t\approx 0$ in the figure. From the discussion in
Sec.~\secsims, however, we recall that the discretization
error is particularly large in this regime. In fact, for the $q=1/100$
model studied in Sec.~\secsims, a second-order Richardson
extrapolation predicts about a $10~\%$ reduction in the amplitude
around the first strong maximum in the $l=2$ waveform which is very
close in magnitude and sign to the deviation of the numerical
from the PP result.
As demonstrated by the upper central panel in Fig.~\ref{fig:waveformsPP},
we find equally good agreement of the numerical $l=2$ multipole with PP
predictions for the less extreme mass ratio $q=1/10$ 
and only a small deviation for the larger mass ratio $q=1/4$ 
(upper top panel in Fig.~\ref{fig:waveformsPP}).
Our findings thus confirm over a wide range of mass ratios the 
observation by Ref.~\cite{Anninos:1994gp}, that there is a weak
dependence of the re-scaled waveforms on the mass ratio.
The $l=3$ mode, on the other hand, is a good discriminator between high- and
low-mass ratios.
This behavior was qualitatively expected, as
higher multipoles are suppressed in the equal-mass case;
by symmetry the $l=3$ mode is absent when the masses are equal.
It is interesting,
however, that even for what one might call a small mass ratio, $q=1/10$,
higher multipoles are still visibly suppressed.
%
\begin{table}
\begin{tabular}{cccccc}
\hline
$q$                                  & 1/1    & 1/4    & 1/10   & 1/100  & PP\\ \hline
$E^{\rm rad}_{\infty}/(M\eta^2)$    &0.00936 & 0.00911 & 0.00985  & 0.0114& 0.0104\\
$v^f_{\infty}$ (km$/$s)             &0.0     &258.0   &250.3  &275.9     & 257.6 \\
\hline
\end{tabular}
\caption{\label{tab:r0dependence} Summary of our results when fitted to
Eqs.~(\ref{r0dep}) and (\ref{eq:fitV}). The last column refers to PP results, as extrapolated 
from Lousto and Price \cite{Lousto:2004pr}. 
}
\end{table}

The total amount of energy radiated in gravitational waves during the
collision depends on the initial separation of the holes. As discussed
in Anninos {\em et al.} \cite{Anninos:1994gp}, two effects contribute
to increasing the
GW energy at larger initial separations; (i) there is more time
to radiate GWs during the infall and (ii) the infalling velocity
is larger. In practice, the second effect is found to be dominant.
Anninos {\em et al.} have accounted for both contributions by defining
\begin{eqnarray}
F_{L}&=&\frac{\int_{L}^{2M}\dot{r}\ddot{r}^2dr}{\int_{\infty}^{2M}
\lim_{L\to\infty}\dot{r}\ddot{r}^2dr}\,,\nonumber\\
\dot{r}&=&\frac{(1-2M/r)\sqrt{2ML/r-2M}}{\sqrt{L-2M}}\,.\nonumber
\end{eqnarray}
One can write the corrections to the radiation emission
\be
E^{\rm rad}_{L}=F_{L}E^{\rm rad}_{\infty}=\left(1-\frac{40M}{9L}
\right)E^{\rm rad}_{\infty}+{\cal O}\left(\frac{M^2}{L^2}\right)\,.\nonumber
\ee
With the above as motivation, we have fitted our results to a
$1/L$ dependence, of the form
\begin{equation}
\frac{E^{\rm rad}(L)}{M\eta^2}=\frac{E^{\rm rad}_{\infty}}{M\eta^2}
\left(1+a_E\,M/L\right )\,,\label{r0dep}
\end{equation}
with $E^{\rm rad}_{\infty}$ the radiated energy for infinite initial
separation.  The results are summarized in Table~\ref{tab:r0dependence}.
We remind the reader that $L$ stands for proper initial separation between
the holes.  We also note that the results in Table \ref{tab:r0dependence}
are normalized by $\eta^2$.  For comparison, we also show in the
last entry of the table the results obtained in the PP limit, within
a linearized calculation. This study was done by Lousto and Price
\cite{Lousto:2004pr} using the same type of initial data; we have used
their Table I to obtain the behavior shown in Table~\ref{tab:r0dependence}
above. We note that already for $q=1/10$ and $q=1/100$ our results are in good
agreement with PP calculations. We remind the reader, however, that
in the $q=1/10$ case there is a larger deviation in the $l=3$ modes.

With the extrapolation above one gets an estimate for the total radiation
of two black holes merging from infinite initial separation. A best fit
of this number as function of mass ratio yields
\begin{equation}
\label{eq:fitA}
\frac{E^{\rm rad}_{\infty}}{M\eta^2}= 0.0110 -0.0088 \eta \, 
\end{equation}
In the PP limit, when $\eta \to 0$, this agrees with the classical PP
calculation, Eq.~(\ref{DRPP}) to within $6\%$, so within the numerical
uncertainties. Overall, the results in Table \ref{tab:models} demonstrate
that we are able to accurately measure amounts of order $E^{\rm rad}\sim
10^{-6}M$ in these fully nonlinear evolutions.


The amount of spurious radiation in the initial data is also
consistent with predictions from linearized gravity. Lousto and Price
performed a detailed analysis of the amount of spurious radiation in the
infall of PPs into massive black holes, using the same type of initial
data \cite{Lousto:2004pr}. Using their Table I for $L>11$,
we find that the amount of spurious radiation varies with $L$ according
to $E_{\rm rad}/(M\eta^2) \sim 0.15(L/M)^{-2.5}$.
For $q=1/100$, for instance, we obtain
$E_{\rm rad}/(M\eta^2)=0.26(L/M)^{-2.55}$.
Thus, we find good agreement in the decay power (roughly $-2.5$) and
also in the proportionality coefficient.

If two BHs with different masses collide head-on, the remnant BH will recoil 
with respect to the center-of-mass frame, due to the emission of energy and 
momentum carried by gravitational waves. Based on PN tools, we have fit
our results to \cite{Fitchett:1983}
\begin{equation}
  \label{eq:fitV}
  v_{\rm recoil} =v^f_{\infty}\, \frac{q^2(1-q)}{(1+q)^5}
    \left(1+b_E\,M/L\right)\,,
\end{equation}
where $v^f_{\infty}$ is a normalized recoil velocity for infinite initial
separation.  The normalized recoil velocity $v^f_{\infty}$ is shown in
Table \ref{tab:r0dependence}.  The point particle limit was considered in
Ref.~\cite{Nakamura:1983hk}, who obtained $v^f_{\infty}=263 {\rm km/s}$ \footnote{note the slight disagreement with the extrapolation of Lousto
and Price's results, shown in Table \ref{tab:r0dependence}}.  We note this
is not a trivial agreement: unlike energy calculations, momentum involves
interference with higher (typically highly suppressed) multipoles.
Overall, our results agree well in the limit of small mass-ratios with the
point particle limit. It is interesting to note in this context that for
both, radiated energy and linear momentum, the numerical results exceed
those obtained from the point particle limit by about $6~\%$. This value
agrees in sign and magnitude with the discretization error obtained for
the $q=1/100$ simulation in Sec.~\ref{sec:sims}. We therefore consider
the discretization error the dominant source of the remaining
discrepancies.

\vspace{0.2cm}
\section{\label{concl} Conclusions}

The simulation of dynamical, interacting black holes has a tremendous
potential to provide answers to some of the most fundamental questions
in physics. Recent developments in experimental and theoretical
physics make this a pressing issue. We refer, in particular, to the
prominent role of BHs in the gauge-gravity duality, in TeV-scale
gravity or even on their own as solutions of the field equations
\cite{Zilhao:2010sr}. Recent work along these lines includes the
successful simulation and understanding of the collision of two
BHs at close to the speed of light in four-dimensional spacetime
\cite{Sperhake:2008ga,Sperhake:2009jz,Sperhake:2010uv,Shibata:2008rq},
the low energy collisions in higher spacetime dimensions
\cite{Zilhao:2010sr,Witek:2010xi,Witek:2010az}, BH scattering in five
dimensions \cite{Shibata:2011fj}, stability studies in
higher dimensions \cite{Shibata:2009ad, Shibata:2010wz,Lehner:2010pn} and
BH evolutions in non asymptotically flat spacetimes \cite{Witek:2010qc}
(for the formalism extension, we refer the reader to Refs.
\cite{Yoshino:2009xp,Zilhao:2010sr,Witek:2010xi,Sorkin:2009bc,Sorkin:2009wh,
Dennison:2010wd}).

We have shown here that Numerical Relativity is capable of simulating
dynamical black holes close to the regime of validity of linear
calculations, and to make contact with
approximation techniques. For this purpose we have evolved head-on
collisions of non-spinning black-hole binaries over a range of
mass ratios from $q=1$ to $q=1/100$.
We obtain radiated energies decreasing from about $5.5 \times 10^{-4}$
for $q=1$ to $10^{-6}$ for $q=1/100$. The recoil reaches
a maximum of about $4~{\rm km/s}$ near $q=3$ and decreases towards
$26~{\rm m/s}$ for $q=1/100$. In the limit of small mass ratios and
extrapolating our results to infinite initial separation, we find the
numerical values for radiated energy and linear momentum to be
$\approx 6~\%$ larger than the point-particle predictions. This
discrepancy agrees rather well in sign and magnitude with the
discretization error obtained from a convergence study of our
$q=1/100$ simulations. It thus appears likely that a significant
part of the remaining differences can be attributed to the
discretization error which mirrors the computational demands
of numerical black-hole binary simulations with such small
mass ratios.

With regard to the waveforms, the most remarkable result is the
suppression of odd $l$ multipoles. While we observe good
agreement between numerical and point-particle results for the
$l=2$ mode, already for $q=1/10$, the numerically calculated
$l=3$ multipole is visibly
suppressed for this case and only agrees well with the PP limit
for $q=1/100$.

Overall, the good agreement for waveforms and radiated energy and
momenta for the case $q=1/100$ demonstrates that numerical techniques
are capable of bridging the gap between linear analysis and the
fully non-linear regime of general relativity.

\vspace{0.2cm}
\begin{acknowledgments}
  We thank Ermis Mitsou for sharing data from his earlier work
  \cite{Mitsou:2010jv}, which helped assess the accuracy of our own PP
  waveforms and fluxes. U.S. acknowledges support from the Ram\'on y
  Cajal Programme of the Ministry of Education and Science of Spain,
  NSF grants PHY-0601459, PHY-0652995 and the Sherman Fairchild Foundation
  to Caltech.  H.W. is funded by FCT through grant SFRH/BD/46061/2008.
  This work was supported by the {\it DyBHo--256667} ERC Starting
  Grant and by FCT - Portugal through projects PTDC/FIS/098025/2008,
  PTDC/FIS/098032/2008 CTE-AST/098034/2008 and CERN/FP/116341/2010.
  C.D.O. acknowledges support
  from NSF grants OCI-0905046 and AST-0855535. E.S. acknowledges support
  from grants NSF 0721915 (Alpaca) and NSF 0904015 (CIGR).
  This research was supported
  by allocations through the TeraGrid Advanced Support Program under
  grant PHY-090003 and grant PHY-100033,
  the Centro de Supercomputaci{\'o}n de Galicia
  (CESGA) under project numbers ICTS-CESGA-120 and ICTS-CESGA-175 and
  DEISA Extreme Computing Initiative (DECI-6). Computations were performed
  on the TeraGrid clusters NICS Kraken, SDSC Trestles, on CESGA's Finis
  Terrae cluster, on the Milipeia cluster in Coimbra and LRZ in Munich.
\end{acknowledgments}



\bibliographystyle{h-physrev4}


\end{document}